\documentclass[dvips]{article}
\usepackage{icrctc07}

\title{The misleading nature of the leaky box models in cosmic ray
physics}
\shorttitle{Misleading nature of the leaky box models}
\authors{A. Codino$^{1}$, F. Plouin$^{2}$.}
\shortauthors{Codino, Plouin}
\afiliations{$^1$INFN and Dip. di Fisica dell'Universit\'a degli
Studi di Perugia, Via A. Pascoli, 06123 Perugia, Italy\\ 
$^2$ Former CNRS researcher, Ecole Polytechnique, LLR, F-91128 Palaiseau,
France}
\email{antonio.codino@pg.infn.it}

\abstract{Many experimental results around and above the energies
where the solar modulation affects cosmic ion fluxes were
quantified, conceptualized  and debated using leaky box models.
These models exploit the notion of equilibrium between creation and
destruction processes of cosmic ions in an undifferentiated
arbitrary volume representing the Galaxy, ignoring the galactic
magnetic field, the size of the Galaxy, the position of the solar
cavity, the spatial distribution of the sources, the space variation
of the interstellar matter and other pertinent observations. Progress
in the measurements of the quoted observational parameters
rendes obsolete the use of the leaky box models. Specific examples
substantiating the inadequacy of the leaky box models are analyzed
such as the conversion of the boron-to-carbon flux ratio into grammage
and the residence times of cosmic ions in the Galaxy. The unphysical
and misleading nature of the leaky box models is ascertained and
illustrated at very high energy.}

\begin{document}
\maketitle

\section{Introduction}
Many experimental results on the primary cosmic radiation at very
low energy gathered in the last five decades by balloon-borne
instruments and satellite experiments have been expressed using
simple theories of galactic cosmic rays referred to as Leaky Box
Models. Two basic physical quantities, namely the gas column
encountered by Galactic cosmic rays ($grammage$) and the related
residence time ($T$), were believed to correctly interpret a number
of measurements. Ritual fittings of the grammage above an adjustable
energy $E_0$ (or the rigidity $R_0$) based on numerous measurements
have the form:
\begin{equation}
g = G {r}^{-\delta}
\end{equation}
where $G$ is a constant grammage at the rigidity $R_0$, $\delta$ a
constant and $r$ = $R$/$R_0$, with $R$ the ion rigidity. Numerical
values in classical interpolations are, for example: $G$=$10.8$ g/cm$^2$,
$\delta$=$0.6$ and $R_0$=$4$ GV \cite{ref1}, or
$G$=$24.0$ g/cm$^{2}$, $\delta$=$0.65$ and $R_0$=$5.5$ GV
\cite{ref2}.

Grammage is converted into residence time by: $T$ = $g$/$\rho$$v$
where $v$ is the ion velocity and  $\rho$ the matter density. 
The extrapolation at very high energy of the residence time caused
fictitious problems, as pointed out by Hillas \cite{ref3,ref4}.
Figure 1 deliberately shows on a linear scale in energy the B/C
flux ratio in the energy band $10$ - $200$ GeV/u indicating no
compelling empirical evidence for a decrease of the grammage of the
form $r^{-\delta}$ which, on the contrary, is well established in
the region $1 - 4$ GeV/u. 
The recent measurements of the B/C
flux ratio at $700$ GeV/u by the Runjob Collaboration
\cite{ref5}, along with the data shown in figure 1, exclude the
functional form \cite{ref1} above $20$ GeV/u. 
The data in figure 1
and the B/C flux ratio [5] vividly testify to the inadequacy of the
grammage fitting $via$ $r^{-\delta}$ to physical reality in an area
believed in past decades to be the realm of the Leaky Box Models.

\begin{figure}[t]         
\begin{center}
\includegraphics [width=0.48\textwidth]{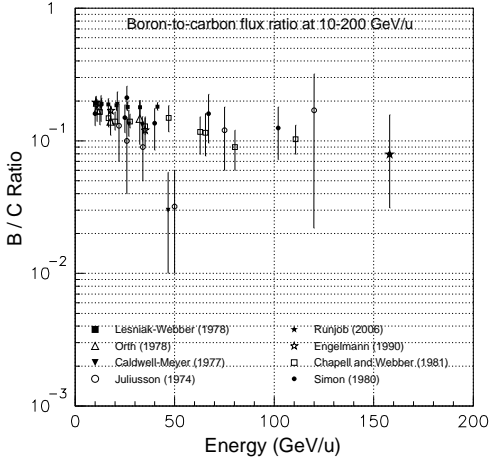}
\end{center}
\caption{Measurements of the boron-to-carbon flux ratios above $10$ GeV/u
up to the energy of $200$ GeV/u displayed in a linear
scale of energy. No decreasing trend of the B/C flux ratio is
evident. Data below $10$ GeV/u may be found elsewhere
\cite{ref18}.}\label{fig1}
\end{figure}

\begin{figure}[t]         
\begin{center}
\includegraphics [width=0.48\textwidth]{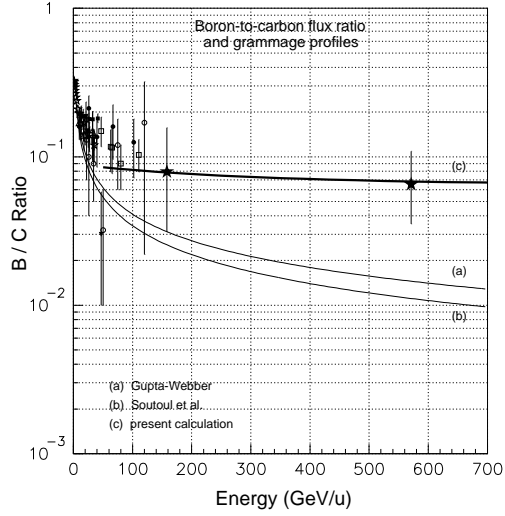}
\end{center}
\caption{Interpolated grammage $via$ $G$ ${r}^{-\delta}$ (thin lines
$a$ and $b$) for two sets of constants mentioned in the
Introduction. It is evident that the slope $\delta$ tuned to the
B/C flux ratio in the range $1-5$ GeV/u is not in accord with
the experimental data above $10$ GeV/u. The computed grammage
(c) is a tiny fraction of that displayed in figure 3.}\label{fig2}
\end{figure}

\section{The unjustified form of the cosmic-ray source power versus energy}

In a suitable Galactic container, at a given energy $E$, two
processes are at work to maintain cosmic-ray intensity at a constant
value: escape from the container and nuclear interactions inside.
The simplified equation describing these processes are:
\begin{equation}
dQ/dE = k (dN/dE) (1/\lambda + 1/f)
\end{equation}
where d$Q$/d$E$ is the source power, d$N$/d$E$ is the differential
intensity in a given location of the disc, $\lambda$ is the nuclear
collision length, $f$ is the average escape length and $k$ a
suitable normalization constant. It is an established result that
the differential intensity at Earth, d$N$/d$E$, measured in many energy
bands, obeys a power law i.e. d$N$/d$E$ = $a$$E^{-\gamma}$ where $\gamma$
is the spectral index and $a$ a constant.

Whenever one of the two terms (1/$\lambda$) and (1/$f$) dominates,
the equation (1) simplifies further. 
At sufficient high energy, especially for light ions,
the term  1/$\lambda$ is negligible compared to 1/$f$.
Using the $ad$ $hoc$ hypothesis that the escape
time has the form $E^{-\delta}$ it follows that d$Q$/d$E$ is
proportional to $E^{-s}$ where $s$= $\gamma$ - $\delta$. 
Therefore from the assumption $E^{-\delta}$ (disproved by the B/C data
above $10$ GeV/u) the unknown form of the source power versus
energy of the cosmic radiation automatically transforms into a power
law with a constant index $s$. The splitting of $\gamma$ in two
constant parts, $s$ and $\delta$, in a large energy band remains
unjustified taking into account, not only the B/C flux ratio at
$160$ and $700$ GeV/u measured by the Runjob experiment
\cite{ref5}, but also the classical data shown in figure 1.

\section{The unphysical grammage extrapolated at high energy}
From simple elaborations of the equation (1) it follows that the
grammage, $g$, is related to the secondary-to-primary flux ratio of
cosmic rays (for example, B/C but also $^{3}$He/$^{4}$He, subFe/Fe,
Nitrogen/Oxygen and others) by the equation:
\begin{equation}
{g \over m}  =  { 1 \over{\sigma_{sp} {N_p \over  N_s}  -\sigma_s}}
\end{equation}
where the $N_s$/$N_p$ is the flux ratio observable by experiments at
a given energy, $m$ the mean mass of the interstellar atom,
$\sigma_p$ is the inelastic nuclear cross sections of the primary
with the interstellar matter and $\sigma_{sp}$ secondary production
rate. Equation (3) applies to a single parent nucleus generating a
unique secondary while primary-to-secondary ratios resulting from
many tributaries have more articulated formulae.

In the limited energy range, $1$ - $5$ GeV/u, there is empirical evidence that
$N_s$/$ N_p$ ratio (i.e. the B/C ratio) is decreasing with
energy. Consequently, since $\sigma_{sp}$ and $\sigma_{p}$ are
smooth functions, the grammage has a decreasing trend with
energy, often interpolated by $G {r}^{-\delta}$ with a constant
$\delta$. On the contrary, in the energy band $10$ - $200$ GeV/u,
the classical measurements  of the B/C flux ratio
\cite{ref6,ref7,ref8,ref9,ref10,ref11,ref12} do not conform to the
same $\delta$ extracted at lower energies, as apparent from the data
shown in figure 2. The functional form ${r}^{-\delta}$ fails just
above $20$ GeV/u, exhibiting a large discrepancy at $700$ GeV/u. 
The B/C flux ratios measured by the Runjob experiment
\cite{ref5} reconfirm the failure of the formula (1) with a unique
$\delta$ in the interval $1$ - $700$ GeV/u.

It is surprising that most of the data in the region $10$ - $100$ GeV/u
shown in figure 1 are simply omitted in the comparison
with the formula $g = G {r}^{-\delta}$ (see, for example, fig.1
\cite{ref20} or fig.1 \cite{ref22}) or with direct calculations
(fig.2 \cite{ref21}).

\begin{figure}[t]         
\begin{center}
\includegraphics [width=0.48\textwidth]{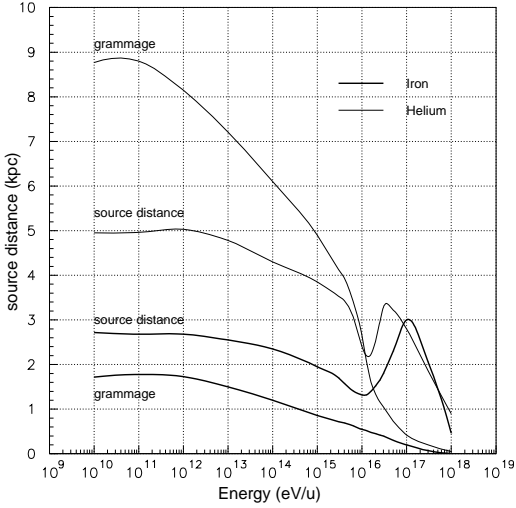}
\end{center}
\caption{The average distance of the sources from the Earth in the
energy range $10^{10}$ - $10^{18}$eV and the related  grammage of
galactic cosmic rays (He and Fe) intercepting the Earth,  expressed
in g/cm$^{2}$ in the same vertical axis. This calculation, unlike
those based on Leaky Box Models, exploits numerous astronomical and
radioastronomical observations (see Section 2 in
\cite{ref14}).}
\label{fig3}
\end{figure}

The $grammage$, the gas column encountered by Galactic cosmic rays,
$g$, can be calculated  using directly the trajectory length $L_D$,
$via$ $g$ = $m$ $n$ $L_D$, where  $n$ the number density of the
interstellar atom. Figure 3 gives the He and Fe grammage versus
energy in the interval $10^{10} - 10^{18}$eV. Details of the
calculation are given elsewhere (13). The computed He grammage
diminishes with a gentle, increasing slope up to $10^{15}$eV
beyond which a steep descend sets on, finally attaining  a minimum
value, at $10^{18}$eV (rectilinear propagation regime). The
dependence of the He and Fe grammage $versus$ energy is explained
elsewhere \cite{ref13,ref14} by the notion of a Galactic basin.

The grammage computed by the direct formula, $g$ = $m$ $n_H$ $L_D$, is
certainly different, by definition,  from that evaluated by the
transport equation used in Leaky Box Models, which relates the
grammage to the B/C flux ratio. For instance, the spatial
distribution of the secondaries should differ from that of the
primaries and this difference is not incorporated in Leaky Box
Models. The carbon grammage versus energy evaluated by the
trajectory length, via $g$ = $m$ $n_H$ $L_D$, is shown in figure 2
(thick line) arbitrarily normalized at $50$ GeV/u to the B/C
ratio of 0.087. The slope of the grammage versus energy in the band
$50$ - $700$ GeV/u is consistent with the data. Note that at lower
energy, below $10$ GeV/u, the grammage for stable $^7Be$ has been
evaluated with the same method resulting in a much steeper slope
\cite{ref15}.

Additional calculations \cite{ref19} of the grammage for the stable
$^9$Be result in a grammage decrease of $30$ per cent in the range
$1$ - $10$ GeV/u both for the spiral and the circular galactic
magnetic fields. Note that the mass and the charge of $^9$Be are
similar to those of B and C, and consequently, the grammage profiles
with energy are analogous. The measured B/C flux ratio decreases
from $0.32$ at $1$ GeV/u to $0.20$ at $10$ GeV/u, exactly
the same decrease of the computed $^9$Be grammage (see fig.2
\cite{ref19}).

\section{The unphysical walls reflecting back cosmic rays in the disc}

The $ad$ $hoc$ assumption that the residence time of cosmic rays is
compatible with a single value (for example, $15 \times 10^6$years)
is unphysical because cosmic rays originated in the Bulge
resides longer that those populating the disc periphery, at $15$ kpc
from the galactic center. The volume where cosmic rays
propagate is not specified and therefore the residence volume is
undefined. The physical motion (migration, diffusion, convection,
trapping or combinations of these classes of displacements) of
cosmic rays is not defined. Since the magnetic field does not exist
in Leaky Box Models and the ion motion is not specified, cosmic rays
should travel freely in the undefined containment volume. But this
free motion, a silent element of any variants of the Leaky Box
Model, implies that a physical, real process, at some boundary of
the disc, reverses the ion motion (reflection). Without the
reflection at some boundary of the disc the grammage cannot
accumulate to high values, because the free traversal of the disc
without reflection entails a grammage of a few milligrams per
cm$^{2}$, some 4 orders of magnitude below the standard $10$ g/cm$^{2}$.
What is the physical mechanism accomplishing this
operation? To date (2007), it remains unknown.

Notice further that the matter density in the disc, adopted in Leaky
Box Models for intrinsic calculation procedure to determine nuclear
spallation rates, is also inconsistent with the matter density
necessary to determine residence time of cosmic rays using
radioactive clock measurements ($^{10}$Be/Be and others). A tangible
sign of this embarassing feature is that the mean gas density in the
disc turns out to be  in the range $0.25 - 0.35$ atom/cm$^{3}$ (see, for
example, \cite{ref16} for the data and \cite{ref17} for the
calculation procedure), a factor $3$-$4$ below the average observed
value of $1$ atom/cm$^{3}$.

A paradox is encountered by extrapolating the residence time at
energies above $10^{17}$ adopting  $G {r}^{-\delta}$ with
$\delta$=$0.65$ \cite{ref2}. The high energy galactic sources would
populate a small volume in the disc, concentric to the Earth, and
they all would reside in the solar system at energies above
$10^{23}$eV.

\bibliography{icrc0705}
\bibliographystyle{plain}

\end{document}